\pdfoutput=1
\documentclass[12pt,a4paper]{article}

\usepackage{ifthen} 
\newboolean{pdflatex}
\setboolean{pdflatex}{true} 

\newboolean{articletitles}
\setboolean{articletitles}{true} 

\newboolean{uprightparticles}
\setboolean{uprightparticles}{false} 

\newboolean{inbibliography}
\setboolean{inbibliography}{true} 

\def\paperauthors{LHCb collaboration}
\def\paperasciititle{Measurement of the prompt-production cross-section ratio sigma(chic1)/sigma(chic2) in pPb collision at 8.16 TeV}
\def\papertitle{Measurement of the~prompt-production cross-section ratio $\sigma(\chi_{c2})/\sigma(\chi_{c1})$ in $p\mkern 1mu\mathrm{Pb}$ collisions at $\sqrt{s_{NN}}$ = 8.16 TeV}
\def\paperkeywords{{High Energy Physics}, {LHCb}} 
\def\papercopyright{\the\year\ CERN for the benefit of the LHCb collaboration} 
\def\paperlicence{CC BY 4.0 licence}
\def\paperlicenceurl{https://creativecommons.org/licenses/by/4.0/}

\usepackage{booktabs}
\usepackage{multirow}

\usepackage[top=1in, bottom=1.25in, left=1in, right=1in]{geometry}

\usepackage{microtype}
\usepackage{lineno}  
\usepackage{xspace} 
\usepackage{caption} 

\usepackage{graphicx}  
\usepackage{color}
\usepackage{colortbl}
\graphicspath{{./figs/}} 

\usepackage{amsmath} 
\usepackage{amssymb}
\usepackage{amsfonts}
\usepackage{upgreek} 

\newcommand*\patchAmsMathEnvironmentForLineno[1]{%
\expandafter\let\csname old#1\expandafter\endcsname\csname #1\endcsname
\expandafter\let\csname oldend#1\expandafter\endcsname\csname
end#1\endcsname
 \renewenvironment{#1}%
   {\linenomath\csname old#1\endcsname}%
   {\csname oldend#1\endcsname\endlinenomath}%
}
\newcommand*\patchBothAmsMathEnvironmentsForLineno[1]{%
  \patchAmsMathEnvironmentForLineno{#1}%
  \patchAmsMathEnvironmentForLineno{#1*}%
}
\AtBeginDocument{%
\patchBothAmsMathEnvironmentsForLineno{equation}%
\patchBothAmsMathEnvironmentsForLineno{align}%
\patchBothAmsMathEnvironmentsForLineno{flalign}%
\patchBothAmsMathEnvironmentsForLineno{alignat}%
\patchBothAmsMathEnvironmentsForLineno{gather}%
\patchBothAmsMathEnvironmentsForLineno{multline}%
\patchBothAmsMathEnvironmentsForLineno{eqnarray}%
}

\usepackage{hyperxmp}

\usepackage[pdftex,
            pdfauthor={\paperauthors},
            pdftitle={\paperasciititle},
            pdfkeywords={\paperkeywords},
            pdfcopyright={Copyright (C) \papercopyright},
            pdflicenseurl={\paperlicenceurl}]{hyperref}

\usepackage[colorinlistoftodos,textsize=scriptsize]{todonotes}

\usepackage[bottom,flushmargin,hang,multiple]{footmisc}

\usepackage[all]{hypcap} 

\usepackage{xspace} 
\usepackage{upgreek}

\def\lhcb   {\mbox{LHCb}\xspace}

\def\MagUp {\mbox{\em Mag\kern -0.05em Up}\xspace}

\ifthenelse{\boolean{uprightparticles}}%
{
 
 \def\Pgamma      {\ensuremath{\upgamma}\xspace}

 \def\Pmu         {\ensuremath{\upmu}\xspace}

 \def\Pchi        {\ensuremath{\upchi}\xspace}                 
 \def\Ppsi        {\ensuremath{\uppsi}\xspace}

 \def\PDelta      {\ensuremath{\Delta}\xspace}                 
 \def\PXi         {\ensuremath{\Xi}\xspace}                 
 \def\PLambda     {\ensuremath{\Lambda}\xspace}                 
 \def\PSigma      {\ensuremath{\Sigma}\xspace}                 
 \def\POmega      {\ensuremath{\Omega}\xspace}                 
 \def\PUpsilon    {\ensuremath{\Upsilon}\xspace}

 \def\PB      {\ensuremath{\mathrm{B}}\xspace}                 
                  
 \def\PD      {\ensuremath{\mathrm{D}}\xspace}

 \def\PJ      {\ensuremath{\mathrm{J}}\xspace}                 
 \def\PK      {\ensuremath{\mathrm{K}}\xspace}

 \def\Pb      {\ensuremath{\mathrm{b}}\xspace}                 
 \def\Pc      {\ensuremath{\mathrm{c}}\xspace}                 
                  
 \def\Pe      {\ensuremath{\mathrm{e}}\xspace}

 \def\Pi      {\ensuremath{\mathrm{i}}\xspace}

 \def\Pp      {\ensuremath{\mathrm{p}}\xspace}

 \def\Ps      {\ensuremath{\mathrm{s}}\xspace}

 \def\thebaroffset{0.0em}
}
{
 
 \def\Pgamma      {\ensuremath{\gamma}\xspace}

 \def\Pmu         {\ensuremath{\mu}\xspace}

 \def\Pchi        {\ensuremath{\chi}\xspace}                 
 \def\Ppsi        {\ensuremath{\psi}\xspace}                 
                  
 \mathchardef\PDelta="7101
 \mathchardef\PXi="7104
 \mathchardef\PLambda="7103
 \mathchardef\PSigma="7106
 \mathchardef\POmega="710A
 \mathchardef\PUpsilon="7107
                  
 \def\PB      {\ensuremath{B}\xspace}                 
                  
 \def\PD      {\ensuremath{D}\xspace}

 \def\PJ      {\ensuremath{J}\xspace}                 
 \def\PK      {\ensuremath{K}\xspace}

 \def\Pb      {\ensuremath{b}\xspace}                 
 \def\Pc      {\ensuremath{c}\xspace}                 
                  
 \def\Pe      {\ensuremath{e}\xspace}

 \def\Pi      {\ensuremath{i}\xspace}

 \def\Pp      {\ensuremath{p}\xspace}

 \def\Ps      {\ensuremath{s}\xspace}

 \def\thebaroffset{0.18em}
}
\newcommand{\offsetoverline}[2][\thebaroffset]{\kern #1\overline{\kern -#1 #2}}%

\makeatletter
\ifcase \@ptsize \relax
  \newcommand{\miniscule}{\@setfontsize\miniscule{4}{5}}
\or
  \newcommand{\miniscule}{\@setfontsize\miniscule{5}{6}}
\or
  \newcommand{\miniscule}{\@setfontsize\miniscule{5}{6}}
\fi
\makeatother

\DeclareRobustCommand{\optbar}[1]{\shortstack{{\miniscule (\rule[.5ex]{1.25em}{.18mm})}
  \\ [-.7ex] $#1$}}

\def\epem       {{\ensuremath{\Pe^+\Pe^-}}\xspace}

\def\mumu       {{\ensuremath{\Pmu^+\Pmu^-}}\xspace}

\def\ellell     {\ensuremath{\ell^+ \ell^-}\xspace}

\def\squark    {{\ensuremath{\Ps}}\xspace}

\def\cquark    {{\ensuremath{\Pc}}\xspace}

\def\bquark    {{\ensuremath{\Pb}}\xspace}

\def\KorKbar {\kern \thebaroffset\optbar{\kern -\thebaroffset \PK}{}\xspace}

\def\D       {{\ensuremath{\PD}}\xspace}

\def\DorDbar {\kern \thebaroffset\optbar{\kern -\thebaroffset \PD}\xspace}

\def\Dp      {{\ensuremath{\D^+}}\xspace}
\def\Dm      {{\ensuremath{\D^-}}\xspace}

\def\DpDm    {\ensuremath{\Dp {\kern -0.16em \Dm}}\xspace}

\def\B       {{\ensuremath{\PB}}\xspace}

\def\BorBbar {\kern \thebaroffset\optbar{\kern -\thebaroffset \PB}\xspace}

\def\Bd      {{\ensuremath{\B^0}}\xspace}

\def\BdorBdbar {\kern \thebaroffset\optbar{\kern -\thebaroffset \Bd}\xspace}

\def\Bs      {{\ensuremath{\B^0_\squark}}\xspace}

\def\BsorBsbar {\kern \thebaroffset\optbar{\kern -\thebaroffset \Bs}\xspace}

\def\jpsi     {{\ensuremath{{\PJ\mskip -3mu/\mskip -2mu\Ppsi}}}\xspace}
\def\psitwos  {{\ensuremath{\Ppsi{(2S)}}}\xspace}

\def\chiczero {{\ensuremath{\Pchi_{\cquark 0}}}\xspace}
\def\chicone  {{\ensuremath{\Pchi_{\cquark 1}}}\xspace}
\def\chictwo  {{\ensuremath{\Pchi_{\cquark 2}}}\xspace}
\def\chicJ    {{\ensuremath{\Pchi_{\cquark J}}}\xspace}
\def\Upsilonres  {{\ensuremath{\PUpsilon}}\xspace}
\def\Y#1S{\ensuremath{\PUpsilon{(#1S)}}\xspace}

\def\proton      {{\ensuremath{\Pp}}\xspace}

\def\LorLbar     {\kern \thebaroffset\optbar{\kern -\thebaroffset \PLambda}\xspace}

\newcommand{\decay}[2]{\ensuremath{#1\!\to #2}\xspace} 

\def\to                 {\ensuremath{\rightarrow}\xspace}

\def\AT#1     {\ensuremath{A_{\mathrm{T}}^{#1}}\xspace}           

\def\C#1      {\ensuremath{\mathcal{C}_{#1}}\xspace}                       
\def\Cp#1     {\ensuremath{\mathcal{C}_{#1}^{'}}\xspace}                    
\def\Ceff#1   {\ensuremath{\mathcal{C}_{#1}^{\mathrm{(eff)}}}\xspace}        
\def\Cpeff#1  {\ensuremath{\mathcal{C}_{#1}^{'\mathrm{(eff)}}}\xspace}       
\def\Ope#1    {\ensuremath{\mathcal{O}_{#1}}\xspace}                       
\def\Opep#1   {\ensuremath{\mathcal{O}_{#1}^{'}}\xspace}                    

\newcommand{\nospaceunit}[1]{\ensuremath{\text{#1}}}       
\newcommand{\aunit}[1]{\ensuremath{\text{\,#1}}}       

\newcommand{\tev}{\aunit{Te\kern -0.1em V}\xspace}
\newcommand{\gev}{\aunit{Ge\kern -0.1em V}\xspace}
\newcommand{\mev}{\aunit{Me\kern -0.1em V}\xspace}
\newcommand{\kev}{\aunit{ke\kern -0.1em V}\xspace}
\newcommand{\ev}{\aunit{e\kern -0.1em V}\xspace}
 
\newcommand{\mevc}{\ensuremath{\aunit{Me\kern -0.1em V\!/}c}\xspace}
\newcommand{\gevc}{\ensuremath{\aunit{Ge\kern -0.1em V\!/}c}\xspace}
\newcommand{\mevcc}{\ensuremath{\aunit{Me\kern -0.1em V\!/}c^2}\xspace}
\newcommand{\gevcc}{\ensuremath{\aunit{Ge\kern -0.1em V\!/}c^2}\xspace}

\def\mub{\ensuremath{\,\upmu\nospaceunit{b}}\xspace}

\def\gsim{{~\raise.15em\hbox{$>$}\kern-.85em
          \lower.35em\hbox{$\sim$}~}\xspace}
\def\lsim{{~\raise.15em\hbox{$<$}\kern-.85em
          \lower.35em\hbox{$\sim$}~}\xspace}

\def\sPlot{\mbox{\em sPlot}\xspace}

\def\sqs   {\ensuremath{\protect\sqrt{s}}\xspace}
\def\sqsnn {\ensuremath{\protect\sqrt{s_{\scriptscriptstyle\text{NN}}}}\xspace}
\def\pt         {\ensuremath{p_{\mathrm{T}}}\xspace}

\def\evtgen     {\mbox{\textsc{EvtGen}}\xspace}

\def\geant      {\mbox{\textsc{Geant4}}\xspace}

\def\photos     {\mbox{\textsc{Photos}}\xspace}

\def\pythia     {\mbox{\textsc{Pythia}}\xspace}

\def\tell1  {TELL1\xspace}
\def\ukl1   {UKL1\xspace}

\newcommand{\eg}{\mbox{\itshape e.g.}\xspace}
\newcommand{\ie}{\mbox{\itshape i.e.}\xspace}

\def\mumugamma  {{\ensuremath{\Pmu^+\Pmu^-\Pgamma}}\xspace}
\def\photon     {\Pgamma}
\def\epos   {\mbox{\textsc{Epos}}\xspace}
\def\pPb    {{\ensuremath{p}}Pb\xspace}

\def\pp     {\ensuremath{pp}\xspace}
\def\pap    {\ensuremath{p{\overline p}}\xspace}
\def\Tm     {Tm\xspace}

\def\ystar  {y^*}
\def\chiconetwo  {{\ensuremath{\Pchi_{\cquark 1,2}}}\xspace}

\usepackage{cite} 
\usepackage{mciteplus}

\begin{document}

\renewcommand{\thefootnote}{\fnsymbol{footnote}}
\setcounter{footnote}{1}

\begin{titlepage}
\pagenumbering{roman}

\vspace*{-1.5cm}
\centerline{\large EUROPEAN ORGANIZATION FOR NUCLEAR RESEARCH (CERN)}
\vspace*{1.5cm}
\noindent
\begin{tabular*}{\linewidth}{lc@{\extracolsep{\fill}}r@{\extracolsep{0pt}}}
 & & CERN-EP-2021-024 \\  
 & & LHCb-PAPER-2020-048 \\  
 & & \today{} \\  
 & & \\
\end{tabular*}

\vspace*{3.5cm}

{\normalfont\bfseries\boldmath\huge
\begin{center}
  \papertitle 
\end{center}
}

\vspace*{2.0cm}

\begin{center}
\paperauthors\footnote{Authors are listed at the end of this article.}
\end{center}

\vspace{\fill}

\begin{abstract}
  \noindent
This article reports the first measurement of prompt $\chi_{c1}$ and $\chi_{c2}$ charmonium production in nuclear collisions at Large Hadron Collider energies. The cross-section ratio $\sigma(\chictwo) / \sigma(\chicone)$  is measured in \pPb collisions at \sqsnn~=~8.16~TeV, collected with the LHCb experiment. 
The \chiconetwo states are reconstructed via their decay to
a $\jpsi$ meson, subsequently decaying into a pair of oppositely charged muons, and a photon, which is reconstructed in the calorimeter or via its conversion in the detector material. 
The~cross-section ratio is consistent with unity in the two considered rapidity regions.
Comparison with a corresponding cross-section ratio previously measured by the LHCb collaboration in \pp collisions suggests that $\chi_{c1}$ and $\chi_{c2}$ states are similarly affected by nuclear effects occurring in \pPb collisions.
\end{abstract}

\vspace*{2.0cm}

\begin{center}
  Published in
  Phys.~Rev.~C 103 (2021) 064905
\end{center}

\vspace{\fill}

{\footnotesize 
\centerline{\copyright~\papercopyright. \href{\paperlicenceurl}{\paperlicence}.}}
\vspace*{2mm}

\end{titlepage}


\newpage
\setcounter{page}{2}
\mbox{~}

\renewcommand{\thefootnote}{\arabic{footnote}}
\setcounter{footnote}{0}

\cleardoublepage

\pagestyle{plain} 
\setcounter{page}{1}
\pagenumbering{arabic}

\section{Introduction}
Collisions of protons with nuclei offer opportunities to study the production and interaction of heavy quarks inside the nucleus.  Charm-quark production in hadron collisions is sensitive to the gluon content of colliding hadrons, and can be used to probe modifications of the parton distributions inside the nucleus \cite{Vogt:2004dh}.  While traversing the nucleus, heavy quarks are also subject to energy loss that can lead to the suppression of bound states \cite{Arleo:2012rs}.  Once the heavy-quark pair exits the nucleus, late-stage interactions with co-moving hadrons can disrupt fully formed quarkonium states \cite{Capella:1996va}. Measurements in proton-nucleus collisions also give an experimental baseline for the interpretation of quarkonium suppression in nucleus-nucleus collisions, where color screening in a deconfined quark-gluon plasma is expected to be a dominant effect \cite{Matsui:1986dk}. 
Studies of quarkonium suppression in \pPb collisions revealed that
the excited states, such as the charmonium \psitwos state or the bottomonium $\Y2S$ and $\Y3S$ states, show a different suppression pattern compared to the \jpsi and $\PUpsilon(1{\rm S})$ states (see \cite{LHCB-PAPER-2017-014,LHCb-PAPER-2015-058,Adare:2016psx,LHCb-PAPER-2018-035,Acharya:2020wwy,Aaboud:2017cif,Sirunyan:2018pse} and references therein). Such a difference cannot be explained by processes taking place during the initial stages of the collision, \ie acting on the quark--antiquark pair. Instead, the processes must occur after the hadronization of the heavy-quark pair into a final state, \eg through dissociation due to interactions with the co-moving matter created at the collision point \cite{Ferreiro:2014bia,Ferreiro:2018wbd}.  Currently, the $\jpsi$ and \psitwos mesons are the only charmonium states which have been measured in collisions of protons with nuclei at the Large Hadron Collider (LHC).  

The \chicJ states, with $J=0,1,2$ denoting the total angular momentum, comprise a triplet of orbitally excited 1P charmonia. They are typically studied in collider experiments via their radiative decay \decay{\chicJ}{\jpsi\photon}, with a subsequent decay \decay{\jpsi}{\ellell}, where $\ell$ denotes electron or muon. 
A selection of recent measurements in \pp and \pap collisions can be found in Refs. \cite{Abulencia:2007bra,LHCB-PAPER-2011-019,LHCB-PAPER-2013-028,Chatrchyan:2012ub,Adare:2011vq}.

The binding energies of \chicJ states are significantly smaller than that of the \jpsi state and greater than the binding energy of \psitwos state \cite{Satz:2005hx}. The small difference in the binding energies of \chicone and \chictwo charmonia makes the ratio of their production cross-sections, $\sigma(\chictwo) / \sigma(\chicone)$, a useful tool to study their sensitivity to final-state nuclear effects, which are expected to be similar for both states. The $\chicJ$ states also form an important feed-down contribution to $\jpsi$ production, so measurements of nuclear effects on $\chicJ$ states can clarify interpretation of the $\jpsi$ data.   Moreover, various efficiency factors and sources of uncertainty cancel out in the ratio, allowing for a more precise measurement.
In nuclear collisions, the 
\chicJ states have been measured by the HERA-B \cite{Abt:2008ed} and PHENIX collaborations \cite{Adare:2013ezl}. To date, no measurement has been reported at the LHC energies. 
 
Here we present the first measurement of the cross-section ratio of promptly produced $\chictwo$ and $\chicone$ states, $\sigma(\chictwo) / \sigma(\chicone)$, in nuclear collisions at the LHC.
The measurement is performed using data collected by the LHCb collaboration in \pPb collisions, at the center-of-mass energy per nucleon pair \sqsnn~=~8.16~TeV,  in 2016.

\section{Experimental apparatus}
The \lhcb detector~\cite{LHCb-DP-2008-001,LHCb-DP-2014-002} is a single-arm forward spectrometer covering the pseudorapidity range $2 < \eta < 5$, designed for the study of particles containing \bquark or \cquark quarks. The detector consists of a high-precision silicon-strip vertex locator (VELO) surrounding the interaction region, a set of four planar tracking stations coupled to a dipole magnet with a 4 \Tm bending power, a pair of ring-imaging Cherenkov detectors to discriminate between different types of charged hadrons, followed by calorimetric and muon systems that are of particular importance in this measurement.
The calorimetric system allows for identification of electrons and photons and consists of a scintillating pad detector (SPD), a pre-shower system (PS), an electromagnetic (ECAL) calorimeter, and a hadronic (HCAL) calorimeter. 
The SPD and PS are designed to discriminate between signals from photons and electrons, while ECAL and HCAL provide the energy measurement and identify electromagnetic radiation and neutral hadrons.
Muons are identified by a system composed of alternating layers of iron and multiwire proportional chambers.

The \pPb data were collected with the LHCb experiment in two distinct beam configurations. 
In the forward configuration, the particles produced in the direction of the proton beam are measured in a center-of-mass rapidity region $1.5 < \ystar < 4.0$, while in the backward configuration, particles produced in the lead-beam direction are measured at center-of-mass rapidity $-5.0 < \ystar < -2.5$. The forward (backward) data sample corresponds to an integrated luminosity of about $ 14 \mub^{-1}$ ($21 \mub^{-1}$).

\section{Data selection}
The analyzed events are selected by a set of triggers designed to record collisions containing the decay  \decay{\jpsi}{\mumu}. 
The \jpsi candidates are reconstructed from a pair of oppositely charged muons with momentum component transverse to the beam, \pt, larger than $700 \mbox{ \mevc}$, originating from a common vertex and an invariant mass within $\pm 42 ~ \mevcc$  of the known \jpsi mass \cite{PDG2020} (corresponding to three times the dimuon mass resolution). The \jpsi candidates are combined with a photon candidate to form a \chiconetwo candidate. 
Photons used in this analysis are classified in two mutually exclusive types: those that converted in the detector material upstream of the dipole magnet and of which the electron and positron tracks were reconstructed in the tracking system (\textit{converted photons}), or those reconstructed through their energy deposits in the calorimetric system (\textit{calorimetric photons}).
The calorimetric photon sample is about an order of magnitude larger than the converted photon sample but has worse mass resolution.
Converted photons are reconstructed from a pair of oppositely
charged electron candidates and are required to have a transverse momentum $\pt$~$> 600$~\mevc and a good-quality conversion vertex \decay{\photon}{\epem}. 
Calorimetric photons are identified 
using the ratio of their energy deposited in the hadronic and electromagnetic calorimeters
and a pair of likelihood-based  classifiers that discriminate photons from electrons and hadrons \cite{LHCb-DP-2018-001,Beteta:2020gui}. 
Calorimetric photons accepted for analysis are required to have $\pt$~$> 1$~\gevc.
The two measurements discussed here are independent given the different reconstruction between the converted and the calorimetric photons.
The selected \mumugamma combinations, which comprise the \chiconetwo candidates, are required to be reconstructed within the pseudorapidity window $2 < \eta < 4.5$ and in the transverse momentum range of $3 < \pt < 15$ \gevc for the converted and $5 < \pt < 15$ \gevc for the calorimetric candidates. 
In order to select the \chiconetwo candidates produced \textit{promptly} at the primary-collision vertex and to suppress \textit{nonprompt} production from $b$-hadron decays occurring away from the primary vertex, an upper limit is imposed on the pseudo-decay time of the candidates, defined as
\begin{equation}
    t_z = \frac{ \left( z_{\rm decay} - z_{\rm PV} \right) \times M_{\chicone} }{p_z},
    \label{eq:pseudo_propertime}
\end{equation}
where $z_{\rm decay} - z_{\rm PV}$ is the difference between the positions of the reconstructed vertex of the \chiconetwo candidate and the primary proton-nucleus collision vertex along the beam axis, $p_z$ is the longitudinal component of the \chiconetwo candidate momentum and $M_{\chicone}$ is the known mass of the \chicone meson \cite{PDG2020}.
The pseudo-decay time is limited to $t_z < 0.1 \text{ ps}$.
The \chicone and \chictwo candidates originating from decays of short-lived resonances, such as \psitwos produced at the interaction point, are also considered in the analysis.

The effects of the detector acceptance as well as of the reconstruction and selection efficiencies are investigated with simulated events. The \chiconetwo signal is generated in \pythia~\cite{Sjostrand:2007gs} with an \lhcb specific configuration \cite{LHCb-PROC-2010-056}. The \chicone and \chictwo states are generated assuming unpolarized production.
The underlying minimum bias forward and backward \pPb collisions are generated using the \epos event generator configured for the LHC \cite{Pierog:2013ria}.
Unstable particles are decayed via \evtgen \cite{Lange:2001uf}. The \decay{\jpsi}{\mumu} decays are corrected for final-state electromagnetic radiation using \photos~\cite{davidson2015photos}.
The response of the detector to the interactions of the generated particles is implemented using the \geant toolkit~\cite{Allison:2006ve,*Agostinelli:2002hh}; for a detailed description see Ref.~\cite{LHCb-PROC-2011-006}.

\section{Data analysis}
This paper aims at measuring the ratio of the cross sections for  prompt \chicone and \chictwo production. 
The cross-section ratio is defined as 
\begin{equation}
     \frac{ \sigma(\chictwo) }{ \sigma(\chicone) } = 
 \frac{ N_{\chictwo} }{ N_{\chicone} }
 \frac{ \varepsilon_{\chicone} }{ \varepsilon_{\chictwo} }
 \frac{ \mathcal{B}(\decay{\chicone}{\jpsi\photon})}{ \mathcal{B}(\decay{\chictwo}{\jpsi\photon}) }.
    \label{eq:cross-section-ratio_def}
\end{equation}
Here, $N_{\chictwo}$ and $N_{\chicone}$ represent the signal yields of the $\chi_{c2}$ and $\chi_{c1}$ states, respectively, and $\varepsilon_{\chictwo}$ and $\varepsilon_{\chicone}$ denote the  efficiencies to reconstruct and select the corresponding state. 
The branching fractions for the \chiconetwo decays are  $\mathcal{B}(\decay{\chicone}{\jpsi\photon}) = \left( 34.3 \pm 1.0 \right) \%$ and $\mathcal{B}(\decay{\chictwo}{\jpsi\photon}) = \left( 19.0 \pm 0.5 \right) \%$ \cite{PDG2020}.

\begin{figure}[tb]
  \begin{center}
    \includegraphics[width=0.49\linewidth]{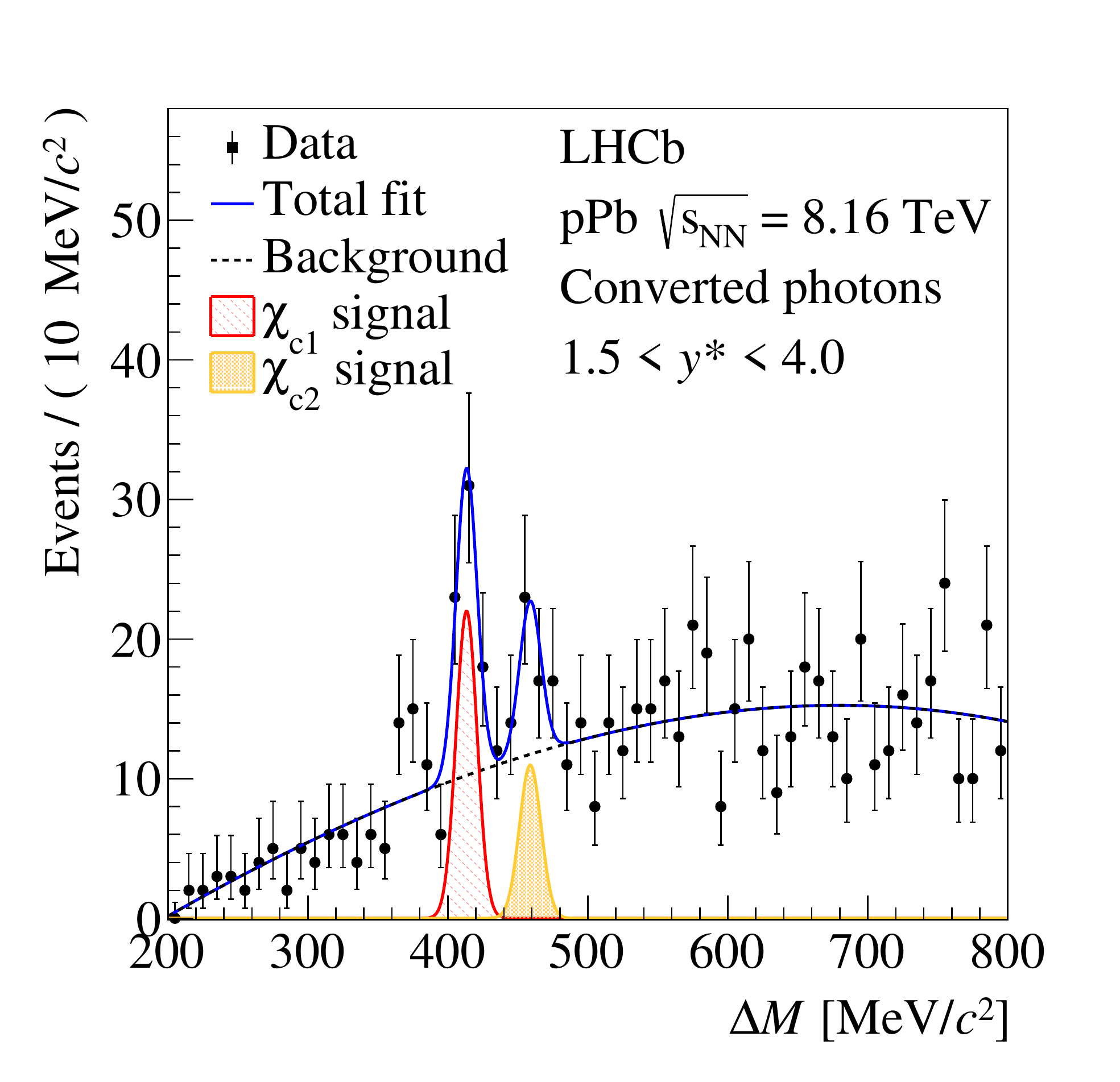}
    \includegraphics[width=0.49\linewidth]{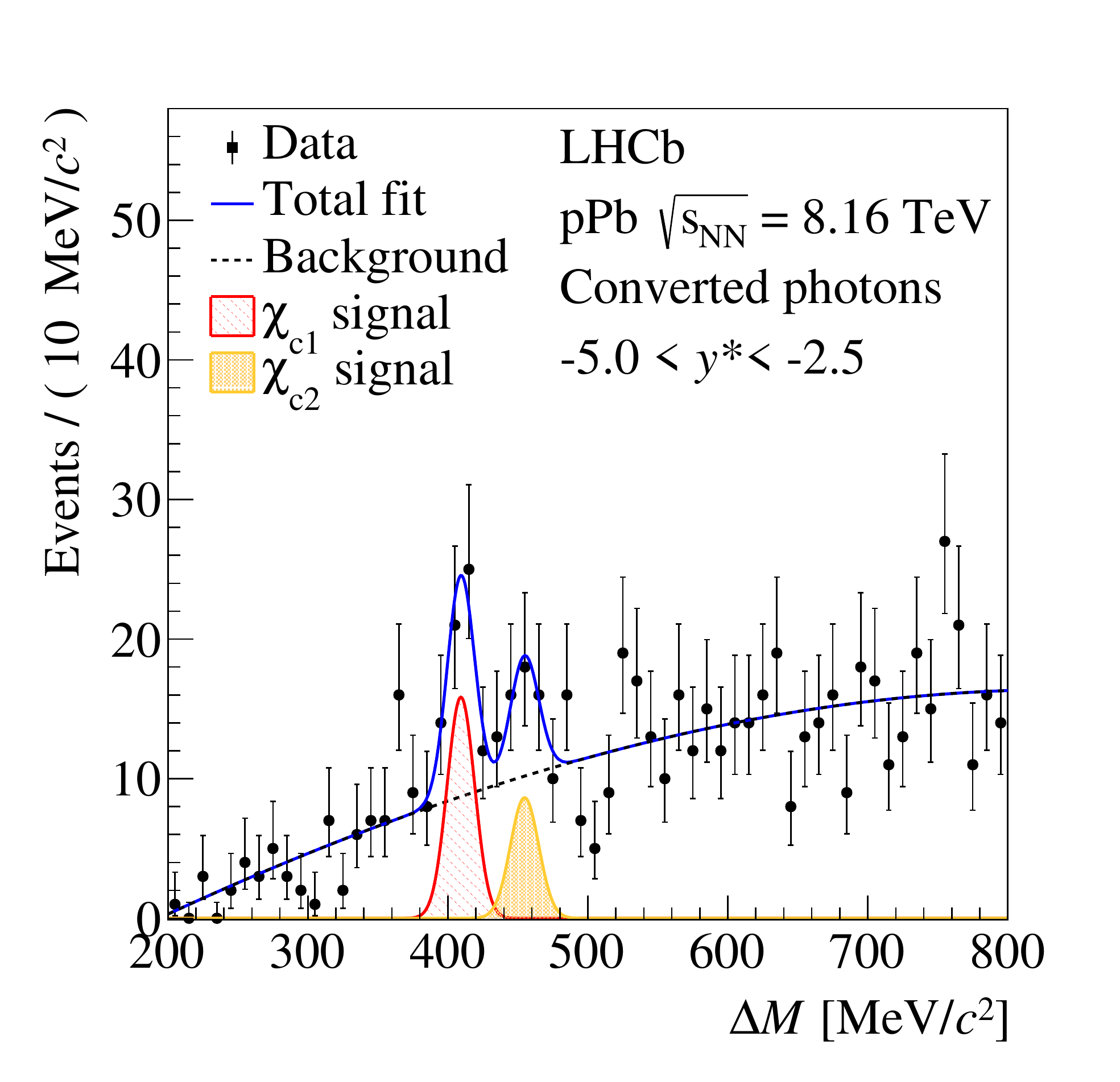}
    \vspace*{-0.5cm}
  \end{center}
  \caption{
    Mass-difference spectra of converted \chiconetwo candidates in forward (left) and backward (right) configuration data. The data are superimposed with a fit (solid blue line) comprising \chicone and \chictwo signals and combinatorial background (dashed black line). 
    }
  \label{fig:signal_extraction_conv}
\end{figure}

\begin{figure}[tb]
  \begin{center}
    \includegraphics[width=0.49\linewidth]{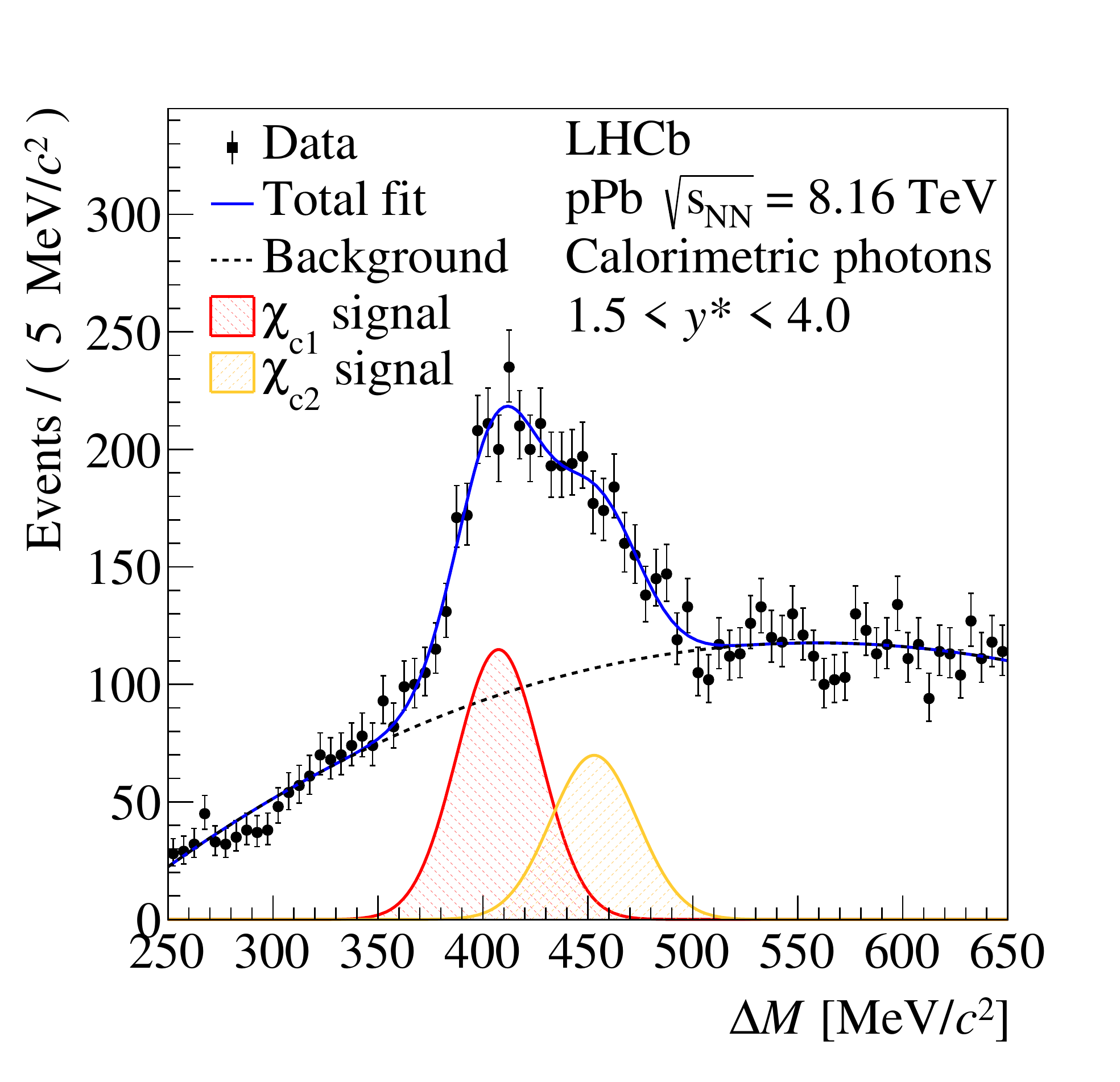}
    \includegraphics[width=0.49\linewidth]{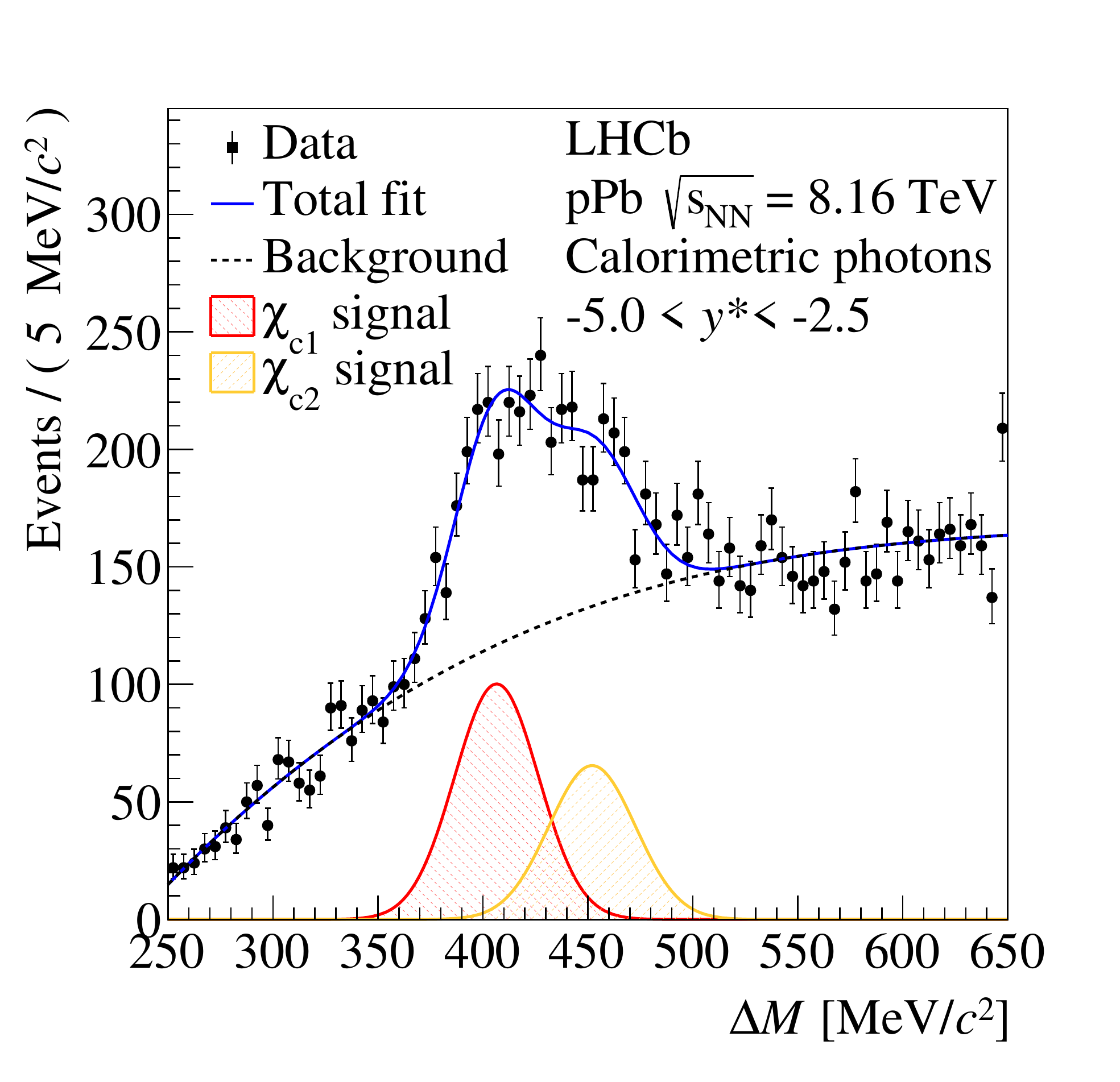}
    \vspace*{-0.5cm}
  \end{center}
  \caption{
    Mass-difference spectra of calorimetric \chiconetwo candidates in forward (left) and backward (right) data. The data are superimposed with a fit result (solid blue line) comprising  $\chicone$ and $\chictwo$ signals and  combinatorial background (dashed black line). 
    }
  \label{fig:signal_extraction_calo}
\end{figure}

The \chicone and \chictwo signal yields are determined by performing a binned maximum-likelihood fit to the spectra of the difference between the invariant mass of the $\mumu \Pgamma$ candidate and that of the \mumu pair, $\Delta M \equiv M(\mumu \gamma) - M(\mumu)$.
The fit function comprises a Gaussian shape for the \chicone and \chictwo resonances and a background component described with a second-order Chebyshev polynomial. 
In the fit, the difference between the values of the $\chi_{c1}$ and $\chi_{c2}$ masses is set to the known mass difference \cite{PDG2020}.  The widths of the $\chi_{c1}$ and $\chi_{c2}$ peaks are set to be equal, following expectations from simulation, and left as a free parameter. The $\chi_{c0}$ peak is also included in the fit, however no significant $\chi_{c0}$ yield is observed.
The fit to the spectra of converted candidates is performed in the range $200 < \Delta M < 800 \ (850)$ MeV/$c^2$ at forward (backward) rapidity. For the calorimetric candidates, the invariant-mass difference spectrum is fitted between $250 < \Delta M < 650$ MeV/$c^2$ in the two rapidity intervals.
The mass-difference spectra of  the converted and calorimetric  samples are shown, together with the fit components, in Figs. \ref{fig:signal_extraction_conv} and  \ref{fig:signal_extraction_calo}, respectively.
In the converted samples, the yield ratio $N_{\chictwo}/N_{\chicone}$ is determined to be $0.51 \pm 0.23$ at forward and $0.56 \pm 0.26$ at backward rapidity, where the uncertainties are statistical. 
In the calorimetric samples, these ratios are  found to be $0.63 \pm 0.08$ at forward and $0.67 \pm 0.10$ at backward rapidity. Individual yields as well as their corresponding significance are listed in Table~\ref{tab:zbytecna_tabulka}.

Since the kinematics of $\chicone$ and $\chictwo$ decays are nearly identical, various detector effects such as tracking and particle-identification efficiencies cancel out in the ratio, so that the efficiency ratio in Eq.~(\ref{eq:cross-section-ratio_def}) can be expressed as 
$\displaystyle{\frac{ \varepsilon_{\chicone} }{ \varepsilon_{\chictwo} } = \frac{ \varepsilon^{\rm acc}_{\chicone} }{ \varepsilon^{\rm acc}_{\chictwo} } \frac{ \varepsilon^{\rm reco}_{\chicone} }{ \varepsilon^{\rm reco}_{\chictwo} }}$.
The factor $\varepsilon^{\rm acc}$ expresses the geometrical acceptance of the decay products to fall within the LHCb acceptance, 
while the factor $\varepsilon^{\rm reco}$ represents the efficiency of selection and reconstruction of the signal candidates.
These correction factors are computed from dedicated simulated events.

\begin{table}[t]
  \caption{
    Yields of \chicone and \chictwo signals with statistical uncertainties and  corresponding  significance (given in standard deviations).
    }
\begin{center}\begin{tabular}{@{} c c c c c c @{} } 
    \toprule
	 \multicolumn{2}{c}{
	 Data sample}	& $N_{\chicone}$    &   Significance &   $N_{\chictwo}$ &   Significance \\
	\midrule
    \multirow{2}{2.3cm}{\centering Converted photons}	& $1.5 < \ystar < 4.0$ & $41 \pm 9$  &  $6.0$ & $21 \pm 8$  & $3.1$	 \\
        &   $-5.0 < \ystar < -2.5$  & $38 \pm 9$	& $4.4$ & $21 \pm 8$ & $3.0$\\
	\midrule
    \multirow{2}{2.3cm}{\centering Calorimetric photons}    & $1.5 < \ystar < 4.0$ & $1151 \pm 69$  &  $15.7$ & $721 \pm 76$  & $9.8$	 \\
        &   $-5.0 < \ystar < -2.5$  & $1004 \pm 73$	& $13.3$ & $676 \pm 82$ & $8.5$\\
	\bottomrule
   \end{tabular}\end{center}
\label{tab:zbytecna_tabulka}
\end{table}

\section{Systematic uncertainties}
The systematic uncertainties on the cross-section ratios are determined as follows.
A systematic uncertainty on the signal extraction is determined by varying the models used in the mass-difference fits.
Several different signal and background models are tested. The signal shapes are varied between Gaussian functions and Voigtian functions (a convolution of a Breit-Wigner and a Gaussian function), and the background shape is varied between second- and third-order Chebyshev polynomials. The natural widths of the \chicone and \chictwo states are narrow compared to the resolution, the Breit-Wigner widths are therefore fixed to the known values~\cite{PDG2020}.
The fit range is varied between $100 \ (150) < \Delta M < 900~\mevcc$ and $200 < \Delta M < 800 \ (850)~\mevcc$ for the converted candidates at forward (backward) rapidity.
For the calorimetric candidates, the fit range is varied between $250 < \Delta M < 650~\mevcc$ and $300 < \Delta M < 600~\mevcc$ in the two rapidity intervals. 
The various choices of signal shape, background parametrization, and range give a total of eight fits to each of the mass-difference spectra in each rapidity interval.
In all cases, the $\chi_{c0}$ peak is also included in the fit; however, no significant $\chi_{c0}$ yield is observed.  
The systematic uncertainty on the yield ratios due to the fitting procedure is assigned as the standard deviation between the values returned by the eight individual fits.
For the converted sample, this systematic uncertainty amounts to $4.9\%$ ($3.2\%$) at forward (backward) rapidity. For the calorimetric sample it is $2.6\%$ ($6.8\%$) at forward (backward) rapidity. The residual background from the nonprompt $\chi_{c1,2}$ production is verified as negligible and shown to cancel out in the ratio, hence no related uncertainty is assigned.
The systematic uncertainty on the acceptance and efficiency corrections includes contributions from the limited size of the simulated samples used to compute the $\varepsilon^{\rm acc}$ and $\varepsilon^{\rm reco}$ factors, and the uncertainty due to the discrepancy of the $\chi_{c1,2}$ and photon properties between data and simulation. 
The latter is estimated using simulated samples,  weighted to reproduce the kinematic distributions of $\chi_{c1,2}$ and photons in background-subtracted data, and obtained using the \sPlot technique, with $\Delta M$ as the discriminating variable~\cite{Pivk:2004ty}. 
The weights are extracted by comparing the transverse momentum and rapidity dependent ratios of the simulated counts $N_{\chicone} / N_{\chictwo}$  with those in data. The simulated $\chicone$ samples are then weighted event-by-event 
and the uncertainty is assessed as the difference between the efficiency ratios computed from simulated samples prior to and after weighting.
In the case of calorimetric photons, an additional weighting process is required in order to recover kinematic distributions of final-state photons observed in the data as well, in a similar event-by-event process as the weights obtained from $\chi_{c1,2}$ kinematic distributions. 
The effect of the photon-identification selection and the reproducibility of relevant variables in simulation are also taken into account. For the converted $\chi_{c1,2}$ sample, the total systematic uncertainty on the acceptance and efficiency equals $9.6\%$ at forward and $14.9\%$ at backward rapidity, while for the calorimetric sample the uncertainty is $8.1\%$ at forward rapidity and $12.4\%$ at backward rapidity. 
The ratio of the branching fractions of the $\chiconetwo \to \jpsi\gamma$ decays contributes with an uncertainty of $3.9\%$.
A summary of contributions to the 
statistical and 
systematic uncertainties of each analyzed sample is given in Table \ref{tab:systematic_uncertainty}.

\begin{table}[t]
  \caption{
    Statistical and systematic uncertainties on the cross-section ratio, $\sigma(\chictwo) / \sigma(\chicone)$. The total systematic uncertainty is also quoted.}
\begin{center}\begin{tabular}{@{} c c c c @{} } 
    \toprule
	 Analyzed sample	& Source	& $1.5 < \ystar < 4.0$	& $-5.0 < \ystar < -2.5$	\\
	\midrule
    \multirow{6}{*}{Converted photons}	& Signal extraction &	$4.9 \%$ 	& $ 3.2 \%$ \\
    & Limited simulation sample size &   $ 5.6 \% $  &   $ 6.5 \% $ \\
	& Efficiency correction	&	$ 7.7 \% $	&	$ 13.4 \% $ \\
	& Branching fraction ratio	&	$3.9\%$ &   $3.9\%$  \\
	\cmidrule{2-4}
	&	Total systematic uncertainty	& $11.4 \%$	& $15.7 \%$ \\
	&   Statistical uncertainty         &	$45.2 \%$ 	& $ 47.0 \%$ \\
	\midrule
    \multirow{6}{*}{Calorimetric photons}    & Signal extraction &	$2.6 \%$ 	& $ 6.8 \%$ \\
	& Limited simulation sample size &   $ 2.5 \% $  &   $ 2.8 \% $ \\
    & Efficiency correction	& $ 7.7 \% $	&	$ 12.1 \% $ \\
    & Branching fraction ratio	&	$3.9\%$ &   $3.9\%$  \\
    \cmidrule{2-4}
	&	Total systematic uncertainty	& $9.3 \%$	& $14.7\%$ \\
	&   Statistical uncertainty         &	$12.2 \%$ 	& $ 14.2 \%$ \\
	\bottomrule
   \end{tabular}\end{center}
\label{tab:systematic_uncertainty}
\end{table}

\section{Results}
\begin{figure}[tb]
  \begin{center}
    \includegraphics[width=0.7\linewidth]{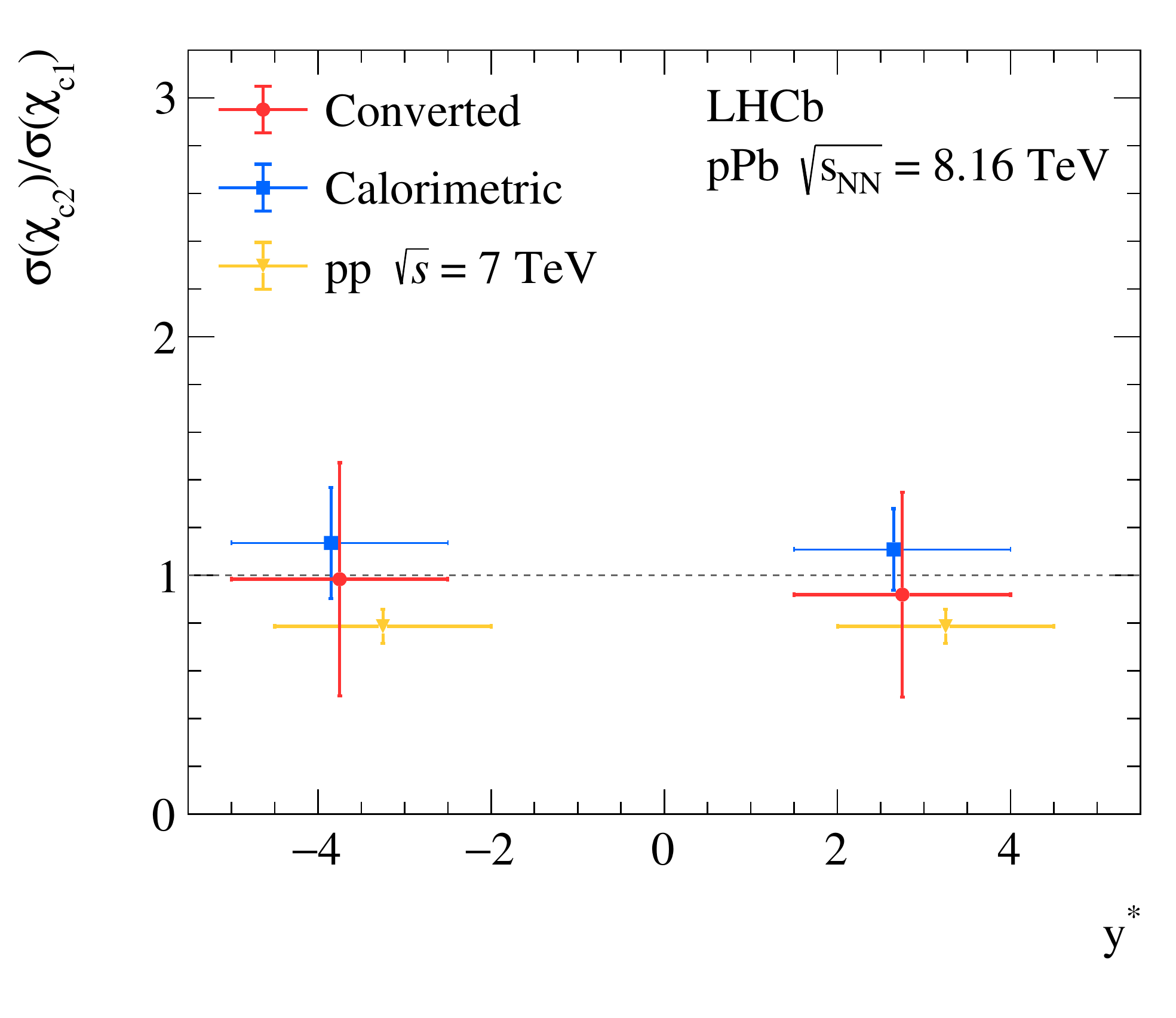} 
    \vspace*{-0.5cm}
  \end{center}
  \caption{
    Cross-section ratio,  $\sigma(\chictwo) / \sigma(\chicone)$ as a function of center-of-mass rapidity $\ystar$, for the $\chictwo$ and $\chicone$ promptly produced in \pPb collisions measured using converted photons (red circles) and calorimetric photons (blue squares). 
    The error bars correspond to the total uncertainties.
    Blue points and vertical uncertainties are shifted horizontally to improve visibility. The \pPb data are compared with results of the converted sample in \pp collisions at \sqs~=~7~TeV \cite{LHCB-PAPER-2013-028} (yellow triangles).
    }
  \label{fig:cross_section_ratio}
\end{figure}

The prompt-production cross-section ratio $\sigma(\chictwo) / \sigma(\chicone)$ in \pPb collisions at the center-of-mass energy per nucleon pair \sqsnn~=~8.16~TeV is shown for the two rapidity regions in Fig. \ref{fig:cross_section_ratio}. 
The ratio measured from converted photons amounts to 
\begin{align*}
    \frac{ \sigma(\chictwo) }{ \sigma(\chicone) } 
    &= 0.92 \pm 0.42 \mbox{ (stat.) } \pm 0.11 \mbox{ (syst.) } && \mbox{ for } 1.5 < \ystar < 4.0, \\
    \frac{ \sigma(\chictwo) }{ \sigma(\chicone) } 
    &= 0.98 \pm 0.46 \mbox{ (stat.) } \pm 0.15 \mbox{ (syst.) } && \mbox{ for } -5.0 < \ystar < -2.5.
\end{align*}
The ratio measured from calorimetric photons is found to be 
\begin{align*}
    \frac{ \sigma(\chictwo) }{ \sigma(\chicone) } 
    &= 1.11 \pm 0.14 \mbox{ (stat.) } \pm 0.10 \mbox{ (syst.) } && \mbox{ for } 1.5 < \ystar < 4.0, \\
    \frac{ \sigma(\chictwo) }{ \sigma(\chicone) } 
    &= 1.14 \pm 0.16 \mbox{ (stat.) } \pm 0.17 \mbox{ (syst.) } && \mbox{ for } -5.0 < \ystar < -2.5. 
\end{align*}
The  cross-section ratios for both converted and calorimetric samples are consistent with unity in both rapidity regions.
The significantly larger yield of the calorimetric sample allows more precise  conclusions on the observed trend to be drawn. 

The cross-section ratio obtained in \pPb data is compared with the corresponding ratio measured in \pp collisions at \sqs~=~7~TeV by the LHCb collaboration \cite{LHCB-PAPER-2013-028}. 
The two measurements are consistent within two standard deviations. While the ratio in the \pp data was measured at a lower center-of-mass energy than that of \pPb collisions, results show that the relative cross-section of different charmonium states is independent of energy at the LHC energy scale \cite{Acharya:2017hjh}. 
Thus, the only aspect to consider in a direct  comparison between the shown \pPb and \pp data is the rapidity range, where the \pPb results are shifted by $-0.5$ in rapidity.
Bearing that in mind, we can express the relative suppression of $\chi_{c2}$ and $\chi_{c1}$ states via the ratio of their nuclear modification factors
\begin{equation}
    \mathcal{R} \equiv \frac{\sigma(\chictwo) / \sigma(\chicone)\scalebox{1.5}{$\vert$}_{p{\rm Pb}} }{\sigma(\chictwo) / \sigma(\chicone)\scalebox{1.5}{$\vert$}_{\pp} }. 
    \label{eq:rppb_ratio}
\end{equation}
Using the more precise calorimetric \pPb results, the ratio of nuclear-modification factors amounts to
$\mathcal{R} = 1.41 \pm 0.21 \mbox{ (stat.) } \pm 0.18 \mbox{ (syst.)}$ at forward and
$\mathcal{R} = 1.44 \pm 0.24 \mbox{ (stat.) } \pm 0.25 \mbox{ (syst.)}$ at backward rapidity,
showing no significant change relative to the \pp ratio in either rapidity region. 
The measured cross-section ratio and ratio of nuclear-modification factors suggest that the nuclear effects have the same impact on both $\chicone$ and $\chictwo$ states within uncertainties, independent of rapidity.

\section{Summary}
In summary, we present the first measurement of $\chi_{c1,2}$ charmonium production in nuclear collisions at the LHC. The cross-section ratio $\sigma(\chictwo) / \sigma(\chicone)$ is consistent with unity for both forward and backward rapidity regions. Moreover, comparison with the ratio measured in \pp collisions hints at a suppression pattern between the two states, which is comparable within uncertainties.  This suggests that the  final-state nuclear effects impact the $\chicone$ and $\chictwo$ states similarly within the achieved precision. 

\section*{Acknowledgements}
%
%
\noindent We express our gratitude to our colleagues in the CERN
accelerator departments for the excellent performance of the LHC. We
thank the technical and administrative staff at the LHCb
institutes.
We acknowledge support from CERN and from the national agencies:
CAPES, CNPq, FAPERJ and FINEP (Brazil); 
MOST and NSFC (China); 
CNRS/IN2P3 (France); 
BMBF, DFG and MPG (Germany); 
INFN (Italy); 
NWO (Netherlands); 
MNiSW and NCN (Poland); 
MEN/IFA (Romania); 
MSHE (Russia); 
MICINN (Spain); 
SNSF and SER (Switzerland); 
NASU (Ukraine); 
STFC (United Kingdom); 
DOE NP and NSF (USA).
We acknowledge the computing resources that are provided by CERN, IN2P3
(France), KIT and DESY (Germany), INFN (Italy), SURF (Netherlands),
PIC (Spain), GridPP (United Kingdom), RRCKI and Yandex
LLC (Russia), CSCS (Switzerland), IFIN-HH (Romania), CBPF (Brazil),
PL-GRID (Poland) and NERSC (USA).
We are indebted to the communities behind the multiple open-source
software packages on which we depend.
Individual groups or members have received support from
ARC and ARDC (Australia);
AvH Foundation (Germany);
EPLANET, Marie Sk\l{}odowska-Curie Actions and ERC (European Union);
A*MIDEX, ANR, Labex P2IO and OCEVU, and R\'{e}gion Auvergne-Rh\^{o}ne-Alpes (France);
Key Research Program of Frontier Sciences of CAS, CAS PIFI, CAS CCEPP, 
Fundamental Research Funds for the Central Universities, 
and Sci. \& Tech. Program of Guangzhou (China);
RFBR, RSF and Yandex LLC (Russia);
GVA, XuntaGal and GENCAT (Spain);
the Leverhulme Trust, the Royal Society
 and UKRI (United Kingdom).

\addcontentsline{toc}{section}{References}
\ifx\mcitethebibliography\mciteundefinedmacro
\PackageError{LHCb.bst}{mciteplus.sty has not been loaded}
{This bibstyle requires the use of the mciteplus package.}\fi
\providecommand{\href}[2]{#2}


\newpage
\centerline
{\large\bf LHCb collaboration}
\begin
{flushleft}
\small
R.~Aaij$^{32}$,
C.~Abell{\'a}n~Beteta$^{50}$,
T.~Ackernley$^{60}$,
B.~Adeva$^{46}$,
M.~Adinolfi$^{54}$,
H.~Afsharnia$^{9}$,
C.A.~Aidala$^{85}$,
S.~Aiola$^{25}$,
Z.~Ajaltouni$^{9}$,
S.~Akar$^{65}$
% ...

{\footnotesize \it

% ...
$^{9}$Universit{\'e} Clermont Auvergne, CNRS/IN2P3, LPC, Clermont-Ferrand, France\\
% ...
$^{25}$INFN Sezione di Milano, Milano, Italy\\
% ...
$^{32}$Nikhef National Institute for Subatomic Physics, Amsterdam, Netherlands\\
% ...
$^{46}$Instituto Galego de F{\'\i}sica de Altas Enerx{\'\i}as (IGFAE), Universidade de Santiago de Compostela, Santiago de Compostela, Spain\\
% ...
$^{50}$Physik-Institut, Universit{\"a}t Z{\"u}rich, Z{\"u}rich, Switzerland\\
% ...
$^{54}$H.H. Wills Physics Laboratory, University of Bristol, Bristol, United Kingdom\\
% ...
$^{60}$Oliver Lodge Laboratory, University of Liverpool, Liverpool, United Kingdom\\
% ...
$^{65}$University of Cincinnati, Cincinnati, OH, United States\\
% ...
$^{68}$Syracuse University, Syracuse, NY, United States\\
% ...
$^{85}$University of Michigan, Ann Arbor, United States, associated to $^{68}$\\
\bigskip
% ...
\medskip
}
\end{flushleft}

\end{document}